\documentclass{PoS}

\def\Journal#1#2#3#4{{#1} {\bf #2}, #3 (#4)}

\def\NPB{{\em Nucl. Phys.} B}

\def\JHEP{{\em J. High Energy Phys.}}

\def\PRD{{\em Phys. Rev.} D}

\def\APPB{{\em Acta Phys. Pol.} B}
\def\IJMPA{{\em Int. J. Mod. Phys.} A}

\def\ra{\rangle}
\def\la{\langle}
\def\dalpha{{\dot{\alpha}}}
\def\dbeta{{\dot{\beta}}}
\def\hph0{{\hphantom{0}}}
\def\C{{\hat{c}}}
\def\r{{\hat{r}}}

\def\overbar{\bar}
\def\blambda{{\overbar{\lambda}}}
\def\bsigma{{\overbar{\sigma}}}
\def\bQ{{\overbar{Q}}}
\def\be{\begin{equation}}
\def\ee{\end{equation}}

\newcommand{\bea}{\begin{eqnarray}}
\newcommand{\eea}{\end{eqnarray}}
\newcommand{\half}{{\scriptstyle{{1\over 2}}}}

\newcommand{\quart}{{\scriptstyle{{1\over 4}}}}
\newcommand{\real}{\relax{\rm I\kern-.18em R}}

\newcommand{\diag}{{\rm diag}}

\newcommand{\veps}{\varepsilon}

\newcommand{\cH}{{\cal H}}

\newcommand{\cG}{{\cal G}}
\newcommand{\cV}{{\cal V}}
\newcommand{\cJ}{{\cal J}}
\newcommand{\cI}{{\cal I}}
\newcommand{\cL}{{\cal L}}
\newcommand{\cS}{{\cal S}}

\newcommand{\Cite}[1]{$\,$\cite{#1}}

\title{The Witten Index Revisited}

\ShortTitle{The Witten Index Revisited}

\author{\speaker{Pierre van Baal}\\
        Instituut-Lorentz for Theoretical Physics\\
        University of Leiden, P.O.Box 9506\\
        NL-2300 RA Leiden, The Netherlands\\
        E-mail: \email{vanbaal@lorentz.leidenuniv.nl}}

\abstract{We attempt to deal with the orbifold singularities in the moduli 
space of flat connections for supersymmetric gauge theories on the torus. 
The fields are restricted to the fundamental domain, containing no gauge
copies, but requiring a boundary condition in field space.}

\FullConference{LIGHT CONE 2008 Relativistic Nuclear and Particle Physics\\
		 July 7-11  2008\\
		 Mulhouse, France}

\begin{document}

\section{Introduction}\label{sec:intro}
We revisit supersymmetric Yang-Mills theories on the torus to study the vacuum 
state in connection with the Witten index\Cite{WiIn}. The torus geometry is 
crucial to preserve the supersymmetry. The index counts the number of 
quantum states (fermionic states with a negative sign). Due to supersymmetry, 
states at non-zero energy occur in fermionic and bosonic pairs, and do not 
contribute to the Witten index. The counting can therefore be reduced to the 
vacuum sector. 

Witten argued that the wave function is constant on the moduli space of flat 
connections and the degeneracy due to the constant abelian gluino components 
costs no energy\Cite{WiIn}. Requiring gauge invariance 
(under Weyl reflections) one has $r+1$ ($=2$ for SU(2)) invariant vacuum 
states which are all bosonic, $|0>$, $U|0>$, $\cdots$, $U^r|0>$, where 
$U=\sum_{a=1}^r \bar\lambda_{\dot\alpha}^a\bar\lambda_{\dot\beta}^a
\epsilon^{\dot\alpha\dot\beta}$.
Untill 1998 there was a puzzle: It didn't agree with the infinite volume gluino 
condensate calculations for SO(N$\geq$7) and the exceptional groups\Cite{ShVa}.
(Additional references are contained in\Cite{Mari}.) Witten solved that by 
finding an extra disconnected component in the moduli space for SO(7), using 
a D-brane orientifold construction\Cite{WiBr}. This means that their are 
isolated points for SO(7) and SO(8) and for SO(N$\ge$9) there is an extra 
component which is like the trivial component for SO(N-7). One therefore finds 
$1+r(SO(N))+1+r(SO(N-7))=h(SO(N))$, or 1+[N/2]+1+[(N-7)/2]=N-2 ($h$ is the 
dual Coxeter number). This is precisely the infinite volume result! 
\begin{table}[h]
\begin{center}
\begin{tabular}{|c | c |cccccc|}
\hline
Group & & &\multicolumn{4}{c}{Vacuum-type} & \\
$G$ & $h$ & 1 & 2 & 3 & 4 & 5 & 6\\
\hline
$SU(N)$    & $N$   & $N$ & & & & & \\
$Sp(N)$    & $N+1$ & $N+1$ & & & & & \\
$SO(2N+1)$ & $2N-1$ & $N+1$ & $N-2$ & & & &  \\
$SO(2N)$   & $2N-2$ & $N+1$ & $N-3$ & & & & \\
$G_2$      & 4  & 3 & 1 & & & & \\
$F_4$      & 9  & 5 & 2 & (1+1) & & & \\
$E_6$      & 12 & 7 & 3 & (1+1) & & & \\
$E_7$      & 18 & 8 & 4 & (2+2) & (1+1) & & \\
$E_8$      & 30 & 9 & 5 & (3+3) & (2+2) & (1+1+1+1) & (1+1) \\
\hline
\end{tabular}
\caption{Contributions to the index for small volumes (A. Keurentjes, 
JHEP {\bf 05}, 001(1999))} 
\end{center}
\vskip-5mm
\end{table}
The field 
theory construction works also for exceptional groups\Cite{KeRS,BoFM}, and is 
based on classifying triples, $\Omega_i=P\exp(\int_0^L A_idx_i)$, which satisfy
$\Omega_i\Omega_j=\Omega_j\Omega_i$. Also here one finds that the index in 
small volumes equals the large volume results.

But a technical problem occurs when using periodic boundary conditions. This
is the breakdown of the adiabatic approximation in the reduction of the 
degrees of freedom to those of the classical vacuum. For SU(2) the classical 
vacuum is defined up to a gauge transformation by the set of zero-momentum 
abelian gauge fields, $A_i=\frac{ic_i^3\tau_3}{2L}$, where $c_i^3$ is constant
and $\vec\tau$ are the Pauli matrices.
Its gauge invariant parametrization is in terms of the Wilson loops that wind 
around the three compact directions of the torus, which are {\em compact} 
variables. This describes the vacuum valley as an orbifold, $T^3/Z_2$ for 
SU(2). Here $g_{[\vec n]}=exp(-2\pi i\vec n\cdot\vec x\tau_3/L)$ (which 
maps $c_i^3$ to $c_i^3+4\pi n_i$) and $g=i\tau_2$ (which maps $c_i^3$ to 
$-c_i^3$) give Gribov copies. The orbifold singularities arise where the 
flat connection is invariant under (part of) the Weyl group (the remnant 
gauge transformations that leave the set of zero-momentum abelian gauge 
fields invariant). For SU(2) their are eight orbifold singularities 
(related to $A=0$ by {\em anti-periodic} gauge transformations), $c_i^3=
-c_i^3+4\pi n_i$, giving $c_i^3=2\pi n_i$ (with $n_i=0$ or $1$). It was 
studied in\Cite{Mari} and we will review some of it here.
 
\section{The Hamiltonian}\label{sec:Ham} 

We choose the dependence on the bare coupling constant such that
\be
\cL=-\frac{1}{4g_0^2}(F^a_{\mu\nu})^2+\frac{i}{2g_0^2}
     \blambda^a\gamma_\mu(D^\mu\lambda)^a.
\ee
The reduction to the zero-momentum degrees of freedom, as in the bosonic case, 
will replace the bare coupling constant by a running and asymptotically free 
coupling constant $g(L)$. The zero-momentum gauge fields are parametrized as 
$A_i=ic_i^a\tau_a/(2L)$. 
The vacuum valley is parametrized by the abelian degrees of freedom. These are 
defined by $r_i$, with $r_ir_j=\sum_a c_i^ac_j^a$, for each $i$ and $j$. 
As said, we may also parametrize the vacuum valley by $r_i=C_i\equiv c_i^3$.

\begin{figure}[htb]
\vspace{5.6cm}
\includegraphics{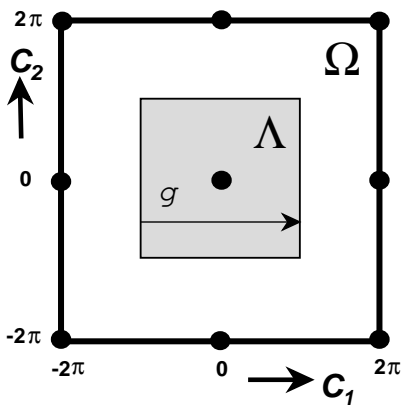}
\caption{A two dimensional slice of the vacuum valley along the $(C_1,C_2)$
plane. The grey square is the fundamental domain $\Lambda$. The dots are 
gauge copies of the origin (which turn out to lie on the Gribov horizon 
$\Omega$, indicated by the fat square).}\label{fig:torus}
\end{figure}

The cell $\vec C\in[-\pi,\pi]^3$ can be used as a fundamental domain $\Lambda$.
Any point on the vacuum valley can be reached by applying suitable gauge 
transformations. This reduces the eight orbifold singularities to one, but 
at a cost: Opposite sides on it's boundary are identified by homotopically 
non-trivial gauge transformations (indicated by {\tt g} in 
Fig.~\ref{fig:torus}). The representations of their homotopy define the 
electric flux quantum numbers as introduced by 't~Hooft\Cite{Tho1}. We will 
here only consider the sector with zero electric flux, i.e. the trivial 
representation, where wave functions at opposite sides are equal. Since the 
orbifold singularity implies massless excitations for the transverse 
fluctuations, we seemingly have no choice: We have to keep the non-abelian 
zero-momentum modes, and thus the adiabatic approximation breaks down.

The energy gap in the fluctuations transverse to the vacuum valley is easily
read off. Close to the origin it is given by $2|\vec C|/L$. Integrating out 
transverse degrees of freedom is only reliable if the energy of the low-lying 
states is smaller than this gap. This energy behaves as $g^{2/3}(L)/L$.
Consider now a sphere of radius $g^{1/3}(L)$ around each orbifold, beyond 
which the adiabatic approximation is accurate. There, the wave function
can be reduced to the vacuum valley and, assuming indeed it has zero energy,
it will be constant. As long as $g(L)$ is small, we may assume the wave 
function in the neighborhood of the orbifold singularity to respect spherical 
symmetry. In the bosonic case, once the wave function spreads out over the 
vacuum valley, any spherical symmetry is quickly lost. In the supersymmetric 
case, however, the reduced wave function on the vacuum valley will become 
constant up to exponential corrections at distances much greater than 
$g^{2/3}(L)$ from the orbifold singularities, which are separated over a
distance $2\pi$. Instead of insisting that the groundstate wave function
is normalizable, we should rather insist on its projection to the vacuum 
valley to become constant. Spherical symmetry near the orbifold singularities 
will dramatically simplify the analysis.

We will now set up the zero-momentum supersymmetric Hamiltonian, and its 
reduction to the gauge invariant, spherically symmetric sector. This is not 
new\Cite{Itoy}, but we will be able to push it to the point where we 
can explicitly construct a complete basis of states that respect these 
symmetries. We start from the supercharge operators,
\be
Q_\alpha=\sigma^j_{\alpha\dbeta}\blambda_a^\dbeta\left(-i\frac{\partial}{
\partial V^j_a}-iB_j^a\right),\quad\bQ_\dalpha=\lambda_a^\beta\sigma^j_{
\beta\dalpha}\left(-i\frac{\partial}{\partial V_j^a}+iB_j^a\right),
\ee
with $V_i^a\equiv c_i^a/(g(L)L)$ and $\sigma^j=\tau^j$ (and $\sigma^0$ the
unit) as $2\times2$ matrices. Restricting to the zero-momentum modes, both the
Weyl spinors $\lambda_a^\beta$ and $\blambda_a^\dbeta$ are constant. Lowering
indices is done with $\epsilon_{\alpha\beta}=\epsilon_{\dalpha\dbeta}=-i
\tau_2$, $\delta_{ab}$ and $\eta_{\mu\nu}=\diag(1,1,1,-1)$ (or $\delta_{ij}$)
for respectively the spinor, group and space-time (or space) indices, and
raising of indices is done with the inverse of these matrices. Repeated
indices are assumed to be summed over, but to keep notations transparent
we will not always balance the positions of the gauge and space indices.
Finally, for zero-momentum gauge fields one has $B_i^a=-\half g\veps_{ijk}
\veps_{abc}V_j^bV_k^c$.

In the Hamiltonian formulation the anti-commutation relations $\{\lambda^{a
\alpha},\lambda^{b\beta}\}=0$, $\{\blambda^{a\dalpha},\blambda^{b\dbeta}\}=0$, 
$\{\lambda^{a\alpha},\blambda^{b\dbeta}\}=\bsigma_0^{\dbeta\alpha}\delta^{ab}$, 
with $\bsigma_0$ the unit $2\times2$ matrix (one has $(\bsigma^\mu)^{\dalpha
\alpha}=\epsilon^{\dalpha\dbeta}\epsilon^{\alpha\beta}(\sigma^\mu)_{\beta
\dbeta}$), give $\{Q_\alpha,\bQ_\dalpha\}=2(\sigma_0)_{\alpha\dalpha}\cH-
2(\sigma^i)_{\alpha\dalpha}V_i^a\cG_a$, where
\be
\cG_a=ig\veps_{abc}\left(V_j^c\frac{\partial}{\partial V_j^b}-
\blambda^b\bsigma_0\lambda^c\right)
\ee
is the generator of infinitesimal gauge transformations, and
$\cH$ is the Hamiltonian density
\be
\cH=-\half\frac{\partial^2}{\partial V_i^a\partial V_i^a}+
\half B_i^aB_i^a-ig\veps_{abc}\blambda^a\bsigma^j\lambda^b V_j^c.
\ee
Splitting the Hamiltonian, $\int d^3x\cH\equiv g^{2/3}(L)H/L$, in its bosonic
and fermionic pieces, $H=H_B+H_f$, we find with $c_i^a=g^{2/3}(L)\C_i^a$
and $\hat B^i_a=-\half\veps^{ijk}\veps_{abd}\hat c_j^b\C_k^d$ that
$H_B=-\half\left(\partial/\partial \C_i^a\right)^2+\half\left(\hat B_i^a
\right)^2$ and $H_f=-i\veps_{abd}\blambda^a\bsigma^i\lambda^b\C_i^d$.

The orbifold singularities, other than at $\C=0$, lie at a distance $b=2\pi 
g^{-2/3}(L)$ in these new variables $\C$ (measured along the vacuum valley 
where $\hat B$ vanishes). We want to solve for the groundstate wave function 
such that for $|\C|\gg 1$ it becomes a constant, after projecting to the 
vacuum valley. As this boundary condition is compatible with spherical
symmetry, i.e. it goes to the same constant for all directions on the vacuum
valley, we will restrict ourselves to wave functions $\Psi(\C)$ that are
spherically symmetric and gauge invariant. We stress again that this is an 
accidental spherical symmetry, that holds in sufficiently small volumes.

\section{Vacuum valley and boundary condition}\label{sec:val}

States with odd fermion number $F$ do not respect the symmetry and particle-hole
duality relates $F=0$ to $F=6$ and $F=2$ to $F=4$\Cite{Mari}. Without fermions
($F=0$) this will reduce to the bosonic Hamiltonian which is known to have 
a non-zero vacuum energy\Cite{Lue1}. We therefore write down the most general 
$F=2$ states, that is the two-spinor states which are symmetric or 
antisymmetric in the gauge 
index (and thus respectively antisymmetric and symmetric in the spinor index),
$|\cV\ra\equiv{\cV_j}^a{\cI^j}_a\equiv-2i\cV_j^c\veps_{abc}\blambda^a_\dalpha
(\bsigma^{j0})^\dalpha_\dbeta\blambda^{b\dbeta}|0\ra$ and $|\cS\ra\equiv
\cS_{ab}\cJ^{ab}\equiv-\cS_{ab}\blambda^a_\dalpha\blambda^b_\dbeta\epsilon^{
\dalpha\dbeta}|0\ra$ (where $\bsigma^{\mu\nu}=\quart(\bsigma^\mu\sigma^\nu-
\bsigma^\nu\sigma^\mu)$, such that $\bsigma^{j0}=\half\tau_j$ as a $2\times2$ 
matrix with the first index up and the second down). Here $\cV_j^a$ and 
$\cS^{ab}=\cS^{ba}$ are arbitrary and covariance allows us to write this as 
${\cV_j}^a=h_1\C_j^a/\r-h_2\hat B_j^a/\r^2+h_3\C_j^b\C_k^b\C_k^a/\r^3$ and 
$\cS^{ab}=h_4\delta^{ab}+h_5\C_j^a\C_j^b/\r^2+h_6\C_j^a\C_j^d\C_k^d\C_k^b/
\r^4$, where $h_i$ depend on $\r^2=(\C_j^a)^2$, $u=\r^{-4}(\hat B_j^a)^2$ and 
$v=\r^{-3}\det\C$. This gives a six dimensional matrix equation for the 
Hamiltonian which can be split into a ``radial" and ``angular" part 
($\hat H_f=H_f/\r$), 
\be
H_{n'n}^{p'p}=\la p',n'|H|p,n\ra=E_p^{\ell_n}\delta_{nn'}\delta^{pp'}+
              \la p',\ell_{n'}|\r^4|p,\ell_n\ra\la n'|\frac{u}{2}|n\ra+
              \la p',\ell_{n'}|\r|p,\ell_n\ra\la n'|\hat H_f|n\ra.
\ee
Our basis explicitly diagonalizes the angular part (``spherical harmonics")
in terms of invariant polynomials.  
\begin{figure}[htb]
\vspace{7.8cm}
\includegraphics{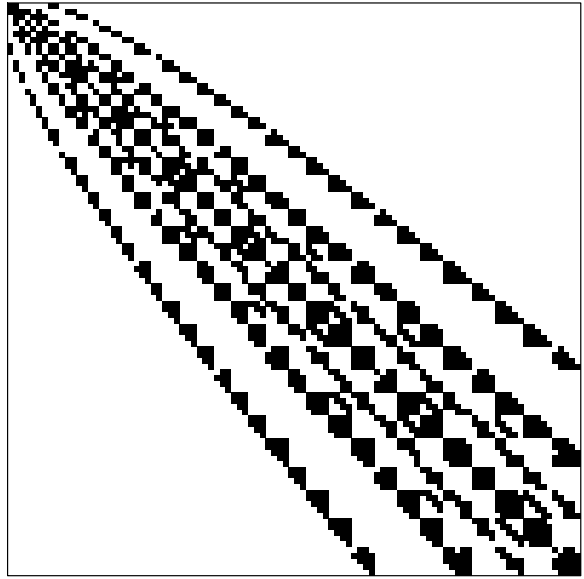}
\caption{Band structure of the reduced Hamiltonian, for the first 100 states.}
\label{fig:band}
\end{figure}
In Fig.~\ref{fig:band} the angular band structure is given by plotting 
$\la n'|u|n\ra\neq0$ or $\la n'|\hat H_f|n\ra\neq0$ as black squares.
The different bands can be traced to come from the selection rules 
$|\delta\ell|=0,2,4$ for the matrix elements of $u$, and $|\delta\ell|=1$ for 
the matrix elements of $\hat H_f$. 

The radial wave function is $\phi_p^\ell(\r)=C_p^\ell\r^{-3}j_\ell(k^\ell_p
\r)$, with $E_p^\ell=\half(k_p^\ell)^2$ and $\ell=2L+3$ ($j_\ell(z)$ is the 
spherical Bessel function and $C_p^\ell$ is the normalization constant). It 
diagonalizes the kinetic term, $-\half\r^{-8}\partial_\r\r^8\partial_\r+
\r^{-2}L(2L+7)=-\half\r^{-3}\left[\r^{-2}\partial_\r\r^2\partial_\r+\half
\r^{-2}\ell(\ell+1)\right]\r^3$. To complete the construction of the basis, 
we need to address the question of boundary conditions (fixing the momenta 
$k_p^\ell$), so that the groundstate wave function becomes a constant, after 
projecting to the vacuum valley. The vacuum valley is characterized by those 
configurations for which $u=0$ (this implies $v=0$), and $\r$ measures the 
(three-dimensional) distance to the origin along this vacuum valley. 

The wave function can be decomposed as $\Psi=\sum_{n=0}^\infty\r^{-3}f_n(\r)
\chi_n(u,v;\r)$, with $\chi_n(u,v;\r)$ normalized eigenfunctions, so that 
$\la\Psi|\Psi'\ra=\sum_{n=0}^\infty\int 4\pi\r^2d\r f_n^*(\r)f^\prime_n(\r)$. 
This means that we have to impose $\partial_\r f_0(\r)=0$ at the boundary of 
the fundamental domain, $\r=b\equiv\pi g^{-2/3}(L)$. This is equivalent to 
$\la\chi_0|\partial_\r(\r^{3}\Psi)\ra=0$, where the inner product is at fixed 
$\r$. That this boundary condition receives corrections is however an 
{\em artifact} of the truncation to the zero-momentum modes. If we take into 
account that in the full theory $\chi_0$ also involves the non-zero momentum 
modes, the (gauge) symmetry guarantees\Cite{Kovb} that {\em at the boundary} 
of the fundamental domain $\partial_\r f_0(b)=0$, and this source of the 
breaking of supersymmetry is absent. Higher order terms in computing the 
effective Hamiltonian are required to deal with this\Cite{Mari}. We therefore 
impose $\partial_\r(\r^{3}\Psi(\r))=0$ at $\r=b\equiv\pi g^{-2/3}(L)$, knowing 
that it is only valid at small $g(L)$. 

\begin{figure}[htb]
\vspace{5.6cm}
\includegraphics{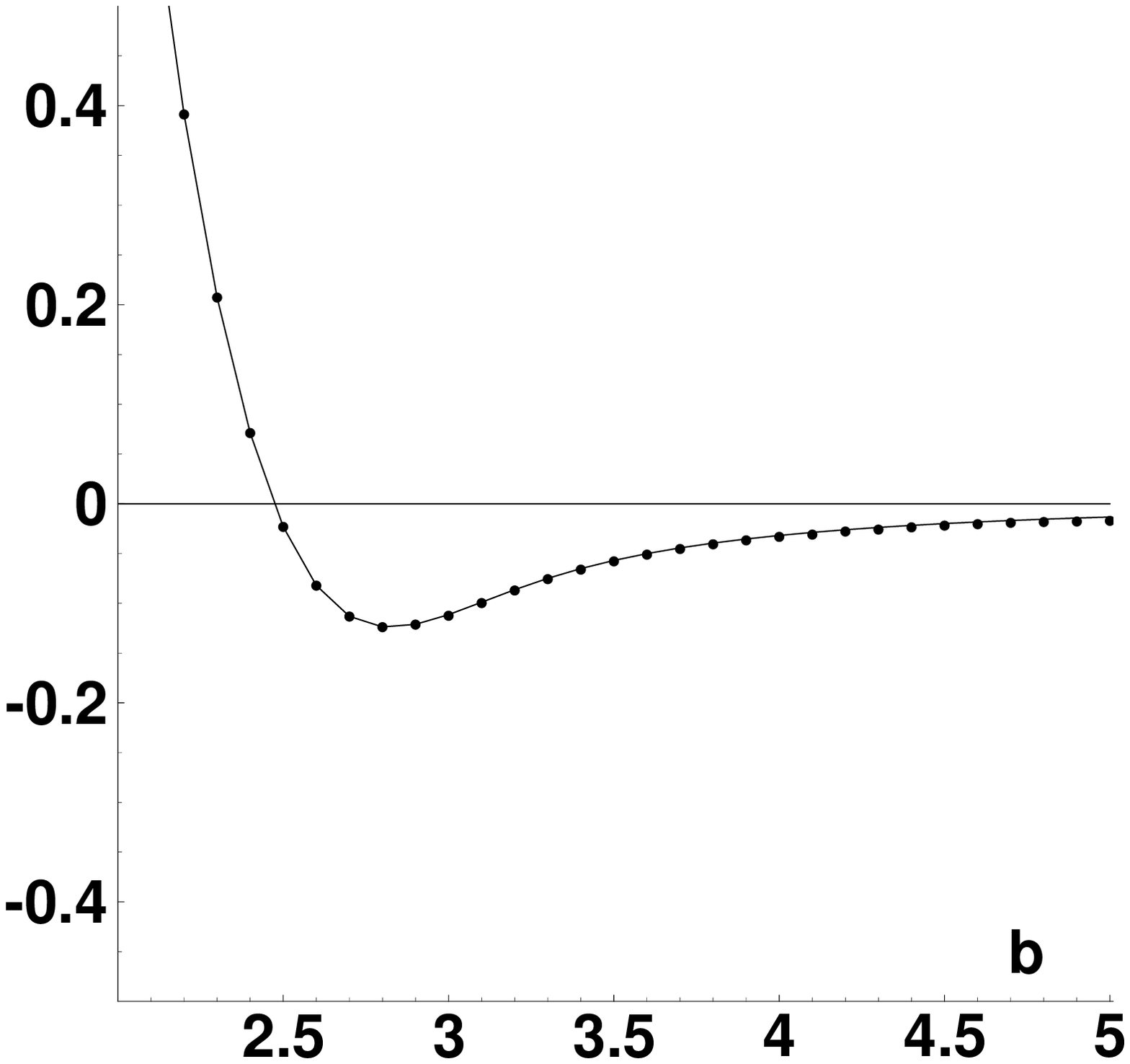}
\includegraphics{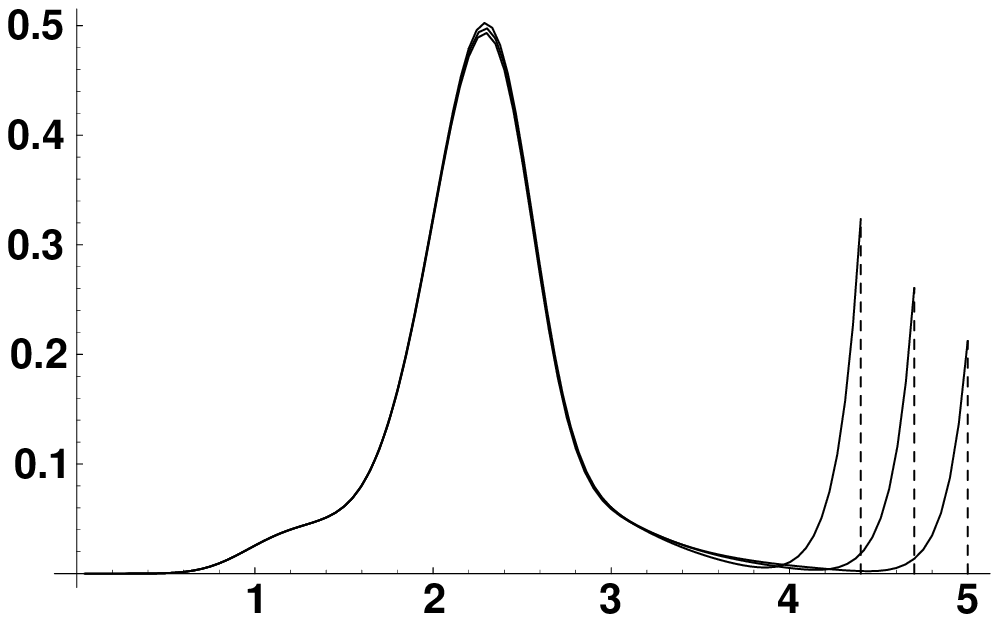}
\vskip-5.2cm \hskip1.5cm ${\bf\times 10^{-4}}$\hskip7cm $E_0$\hfill\vskip.4cm
\hskip1.55cm ${f^2\!\!-\!f_0^2}$ \hfill \vskip2.7cm \hskip4cm 
${\rightarrow\hat r}$ \vskip.5cm
\caption{On the right is shown $f^2(\r)-f_0^2(\r)$ extracted from the 
groundstate wave function $\Psi_0$, satisfying the boundary condition 
$\partial_\r(\r^3\Psi_0)(b)=0$, for $b=4.4$, 4,7 and 5.0. We normalized 
with respect to $b=5$, such that at $\r=2$ all $f^2$ agree. On the left 
is shown the groundstate energy itself, using up to 20 radial modes for 
each of the 420 harmonics together with the lower bound from Temple's 
inequality (indicated by the dots).}\label{fig:Ebs}
\end{figure}

A direct measure for the failure of the Born-Oppenheimer approximation is given 
by $f^2(\r)-f^2_0(\r)\equiv\sum_{n=1}^\infty f_n^2(\r)$, which as expected, 
deviates from zero when the gap is small, but {\em also} when we approach the 
boundary at $\r=b$ (due to the zero-momentum approximation). In 
Fig.~\ref{fig:Ebs} we show the results for $b=4.4$, 4.7 and 5.0, to illustrate 
that this second deviation decreases with increasing $b$ (or decreasing 
coupling). This shows that the mismatch between the boundary condition and the 
truncation of the effective Hamiltonian does not affect the wave function in 
the neighborhood of the orbifold singularities, where the failure of the 
adiabatic approximation is non-perturbative. Finally Fig.~\ref{fig:Ebs} also 
shows the groundstate energy $E_0$ (the curve gives the upper and the dots 
the lower bound). To keep the size of the matrix manageable, the components 
of the eigenvectors are removed when they are in absolute value below a 
threshold, without significantly affecting the accuracy. This process of 
{\em pruning} is performed iteratively, increasing the number of radial modes 
per angular state. Together with Temple's inequality\Cite{Kovb} this is 
extremely efficient to optimize the accuracy and achieve numerical control. 

We have only considered SU(2) to illustrate how to go beyond the adiabatic 
approximation. Our analysis shows that, although the orbifold singularities 
are cause for concern, in the end they do not upset the result for the Witten 
index. For other groups this is much harder. Of course, a numerical analysis 
can never be entirely conclusive in deciding a theoretical issue that involves 
the counting of {\em exact} zero-energy states. The computer code is still 
available at www.lorentz.leidenuniv.nl/vanbaal/susyYM (partly they could 
be used for other studies\Cite{Wosi}). 

\section*{Acknowledgments}
First I wish to thank Jean-Fran\c cois Mathiot for allowing me to speak
on LC2008, particularly because it was on the Witten index. Apparently
I could still present it, and ultimately write it down. Also thanks to 
Koenraad Schalm, who read the manuscript. But I must now also 
thank Margarita Garc\'{\i}a P\'erez and Tony Gonz\'alez-Arroyo for inviting 
me to the Benasque Workshop in 25 February-2 March 2007 and Christian Lang for 
iniviting me to the ``Graduate Day" in Graz on 15 June 2007. Special 
thanks go to Simon Hands (and many other people) for letting me still 
participate in the Newton Institute programme ``Strong Fields, Integrability
and Strings", 19 August-6 October 2007 (while I was there, Bernard Piette 
invited me also for the ``Classical Field Theory Workshop", which took place 
at Durham, 20-21 September 2007). I am also grateful for the invitation to 
the INT program ``From Strings to Things" in Seattle, 5-17 May 2008, and
espesially for the invitation to Galileo Galilei Institute for Theoretical 
Physics in Florence from 26 May-14 June 2008. Valya Zakharov and many other 
people have made this the best workshop I attended since my stroke, and I 
would like to thank especially Mithat \"Ursal for taking me serious. Finally 
I thank Ayse Kizilersu and Tony Williams for inviting me to the ``T(R)OPICAL 
QCD" workshop in Port Douglas, 27 July-1 August 2008. At all of these I gave 
seminars about Calorons, in one way or another.

\end{document}